\documentclass[sigconf, 10pt]{acmart}
\renewcommand\footnotetextcopyrightpermission[1]{} % removes footnote with conference information in first column
\makeatletter
\renewcommand\@formatdoi[1]{\ignorespaces}
\makeatother

\usepackage[english]{babel}
\usepackage{blindtext}
\usepackage{subfigure}
\usepackage{algorithm}
\usepackage{algorithmic}

\acmConference[Conference'XX]{Conference}{XX 20XX}{XX}
\renewcommand\footnotetextcopyrightpermission[1]{} % removes footnote with conference info
\setcopyright{none}
\newcommand{\parabf}[1]{\medskip\noindent\textbf{#1}}

\newcommand{\paraf}[1]{\noindent\textbf{#1}}
\newcommand{\cut}[1]{}

\begin{document}
% \title{A \LaTeX\ Template for SIGCOMM 18}
\title{Is Network the Bottleneck of Distributed Training?}

% CCS 
\begin{CCSXML}
<ccs2012>
   <concept>
       <concept_id>10010147.10010257</concept_id>
       <concept_desc>Computing methodologies~Machine learning</concept_desc>
       <concept_significance>500</concept_significance>
       </concept>
   <concept>
       <concept_id>10003033.10003079.10011704</concept_id>
       <concept_desc>Networks~Network measurement</concept_desc>
       <concept_significance>500</concept_significance>
       </concept>
 </ccs2012>
\end{CCSXML}

\ccsdesc[500]{Networks~Network measurement}
\ccsdesc[500]{Computing methodologies~Machine learning}

% end CCS

%\titlenote{Produces the permission block, and copyright information}
%\subtitle{Extended Abstract}

\author{Zhen Zhang$^1$, Chaokun Chang$^2$, Haibin Lin$^2$, Yida Wang$^2$, Raman Arora$^1$, Xin Jin$^1$}
% \author{Firstname Lastname}
% \authornote{Note}
\affiliation{
   \institution{$^1$Johns Hopkins University, $^2$Amazon Web Services}
%   \streetaddress{Address}
%   \city{City}
%   \state{State}
%   \postcode{Zipcode}
}
% \email{email@domain.com}

\renewcommand{\shortauthors}{Zhen, et al.}

\begin{abstract}
Recently there has been a surge of research on improving the communication
efficiency of distributed training. However, little work has been done to
systematically understand whether the network is the bottleneck and to what extent.

In this paper, we take a first-principles approach to measure and analyze the
network performance of distributed training. As expected, our measurement
confirms that communication is the component that blocks distributed training
from \emph{linear} scale-out. However, contrary to the common belief, we find
that the network is running at low utilization and that if the network can be
\emph{fully} utilized, distributed training can achieve a scaling factor of
\emph{close to one}.
Moreover, while many recent proposals on gradient compression advocate
over 100$\times$ compression ratio, we show that under full network utilization,
there is \emph{no need} for gradient compression in 100 Gbps network.
On the other hand, a lower speed network like 10 Gbps requires only 2$\times$--5$\times$ gradients 
compression ratio to achieve almost linear scale-out.
% Moreover, while recent proposals on application-level
% communication scheduling and gradient compression suggest significant
% improvements, we show that under full network utilization, these solutions can
% improve the performance by at most XX\%.
% Compared to application-level
Compared to application-level techniques like gradient compression,
network-level optimizations do not require changes to applications
and do not hurt the performance of trained models. 
As such, we advocate that the real challenge of distributed training is
for the network community to develop high-performance network transport 
to fully utilize the network capacity and achieve linear scale-out.

\end{abstract}

%%% Local Variables:
%%% mode: latex
%%% TeX-master: "../paper"
%%% End:

\maketitle

\section{Introduction}

Deep Learning is a fundamental building block of modern Internet services, from
personalized recommendation and language translation to content understanding
and voice control. A Deep Neural Network (DNN) model is first trained on a
dataset to achieve high accuracy or other evaluation metrics and then deployed to target platforms to serve requests from end-users. We focus on training in this paper, which is critical to generate high-quality models for deep learning applications.

DNN models are getting larger and deeper. The famous analysis from
OpenAI~\cite{openai-report} shows that the amount of computing needed to train the
state-of-the-art model doubles every 3.4 months, while in comparison, the number
of transistors on a chip only doubles every 18 months even when Moore's law is
still effective. With the end of Moore's law, people have turned to specialized
processors such as GPUs~\cite{GPU-supercomputing} and TPUs~\cite{TPU} to scale up computation. Yet,
compared to the fast-growing demand of DNN models, the computing capability
provided by a single chip is still limited.

As a result, training large DNN models are inevitably getting more and more
distributed by scaling out. The dream for every scale-out system is
\emph{linear} scalability. That is, given that the throughput of a single device is
$T$, the throughput of a system with $n$ devices should be $n T$. Let the
throughput actually achieved by the system with $n$ devices be $T_n$. We define
the scaling factor as
\begin{equation}
\text{scaling~factor} = \frac{T_n}{n T}.
\label{equ:scaling}
\end{equation}
Linear scale-out requires the scaling factor to be 1 for any $n$.

Distributed training with data parallelism strategy includes multiple iterations. Each iteration
can be divided into a computation phase and a communication phase. In the
computation phase, each worker feeds a batch of data into the model, and
performs a forward pass and then a backward pass of the model, to compute the
gradients for learnable parameters. In the communication phase, the workers exchange
their gradients, and compute the average to update the parameters via all-reduce operations.

It is a common belief that the network bandwidth is the bottleneck that prevents distributed training from
scaling linearly. In particular, the computation phase is
embarrassingly parallel, as each worker processes its own batch independently.
The throughput of $n$ workers is $n$ times that of one worker for the
computation phase, and only the communication phase can slow down the training
process.

In response to this, there has been a surge of research from machine learning
and systems communities on improving the communication efficiency of distributed
training in recent years~\cite{peng2019generic, lin2017deep, sapio2019scaling, wen2017terngrad, narayanan2019pipedream, aji-heafield-2017-sparse, alistarh2016qsgd, chen2018adacomp, konevcny2016federated, NIPS2018_7405, wang2018atomo,ivkin2019communication,rothchild2020fetchsgd}. 
% The research can be broadly classified into
% two categories. One is to leverage application-layer information to schedule the
% communication to better overlap it with computation. The other is to exploit
% application-layer knowledge to compress the gradients to reduce the
% communication size.
These works are primarily done at the application layer, assuming that the
network has done its best to maximize  communication efficiency. Yet, little
work has focused on systematically understanding whether the network is the
bottleneck and to what extent.

In this paper, we take a first-principles approach to measure and analyze the
network performance of distributed training. We perform a measurement study on the
training throughput of several representative DNN models on AWS.
%{ In our study,
% we use Horovod~\cite{sergeev2018horovod} to train these models on a set of workers, and vary the
% number of workers and the network speed in the measurement. Horovod is one of
% the most widely-used frameworks for distributed training, being used as the
% baseline in many previous works. }
Our measurements show that the system can achieve a scaling factor of only 60\%
with 64 workers (eight servers with eight GPUs each) for VGG16. As expected, the
measurement confirms that communication is the component that prevents distributed
training from linear scale-out. 
%Specifically, the computation time of each
% batch in a worker does not change, despite the number of workers and the network
% speed, as the computation is embarrassingly parallel
% phase that contributes to the longer per-batch time when the system has more workers.
However, contrary to the common belief, we find that the network bandwidth is not the
bottleneck, because it is running at low utilization. While the network provides
up to 100 Gbps bandwidth for each server, the communication phase uses no
more than 32 Gbps for transferring gradients. We further confirm that the low network utilization is not 
due to the CPU bottleneck.
In fact, the CPU only
runs at 14\%--25\% utilization in the communication phase. 
% Therefore, the root cause is the network stack
% implementation that is not making the best use of the high bandwidth provided by
% modern high-speed networks.

Then the natural question is what if the network can run at 100\% utilization.
We take a white-box approach to get timing information of layer-wise computation in model training.
Based on the logging results, we perform a what-if analysis, in which we control 
the network bandwidth and assume full bandwidth utilization. The results of the analysis 
show that with full network utilization, distributed training can achieve a scaling factor of 
over 99\%. We further extend the what-if analysis with an application-layer optimization---gradient compression. Based on further analysis, 
we find that a compression ratio ranging from 2$\times$ to 5$\times$
is good enough for distributed training to achieve a scaling factor of close to 100\% in 10 Gbps network.

Compared to application-layer optimizations, we argue that network-layer
optimizations should be prioritized for speeding up distributing training. First,
network-layer optimizations are transparent to the applications. They do not
require any changes to the applications or the training systems. Second, unlike
lossy gradient compression in the application layer, network-layer optimizations
do not hurt training convergence rate or model performance.

In conclusion, we make two major contributions. First, we perform a
measurement study to systematically measure and analyze the performance
bottleneck of distributed training. Contrary to the common belief, it unveils
that the network speed is not the problem, but the software implementation of
the communication phase is. Second, we perform a what-if analysis to evaluate
the benefits of high-performance network transport for distributed training.
It reveals that merely optimizing the network transport can already increase the
scaling factor to close to 100\%, and that additional application-layer
optimizations are only required in lower speed networks and we do not need
aggressive optimizing strategies claimed in past works~\cite{seide20141, lin2017deep, lim20183lc}.
As such, we advocate that
the real challenge is for the networking community to develop high-performance
network transport for distributed training to fully utilize the network capacity and
achieve linear scale-out.

\parabf{Open-source.} The code is open-source and available at \href{https://github.com/netx-repo/training-bottleneck}{https://github.com/netx-repo/training-bottleneck}.

%%% Local Variables:
%%% mode: latex
%%% TeX-master: "../paper"
%%% End:

%\input{sections/background}
\section{Profiling Training Performance}

In this section, we describe an empirical study we performed to measure and analyze the
bottleneck in distributed training.

\subsection{Profiling Setup}

\paraf{Training hardware.} The experiments are conducted on Amazon Web Services
(AWS). We use Amazon EC2 p3dn.24xlarge instances
with 8 GPUs (NVIDIA Tesla V100), 96 vCPUs (2.5 GHz Intel Xeon P-8175M
processor), 768 GB main memory, 256 GB GPU memory (32 GB for each GPU), and 100 Gbps network
bandwidth. The 8 GPUs on each instance support NVLink for high-performance
peer-to-peer GPU communication. We vary the number of instances from 2 to 8
(i.e., from 16 GPUs to 64 GPUs) in the experiments to evaluate the scaling factor.

\parabf{Training software.} We use Horovod~\cite{sergeev2018horovod} as the distributed training
framework. Horovod is one of the most widely-used frameworks for distributed
training. It supports popular deep learning frameworks such as
TensorFlow~\cite{tensorflow}, PyTorch~\cite{pytorch} and MXNet~\cite{mxnet}. It uses the all-reduce
strategy for distributed training, which performs an all-reduce operation among
all workers after each iteration to compute the average of the gradients for
parameter update. Horovod uses a combination of NCCL and MPI as the underlying
 layer to implement all-reduce. We use PyTorch as
the training framework for a single GPU, and use Horovod to scale it to multiple
GPUs. The software versions used in the experiments are Horovod 0.18.2, PyTorch
1.3.0, Torchvision 0.4.1, NCCL 2.4.8, cuDNN 6.6.0.64, and Open MPI 4.0.2. Horovod, NCCL, and Open MPI use Linux kernel TCP. 

\parabf{Workloads.} We use three models in the experiments, i.e.,
ResNet50~\cite{resnet}, ResNet101~\cite{resnet} and VGG16~\cite{simonyan2014very}. We choose these models
because they are widely used in computer vision 
and distributed training benchmarks.
Also, they have representative characteristics.
Specifically, ResNet50, ResNet101, and VGG16 have small, medium, , and large parameter sizes, respectively. The model sizes are 97 MB for ResNet50,
170 MB for ResNet101, and 527 MB for VGG16. Besides, VGG16 has a layer
with 400MB parameters, while parameters in ResNet series are distributed more
evenly.
ImageNet dataset~\cite{deng2009imagenet} is used for experiments, and 
we fixed the batch size to 32 for each worker involved in training. 
% Batch size, 32, is fixed for each worker, not matter how many workers

\begin{figure}[t]
    \centering
    \includegraphics[width=0.9\linewidth]{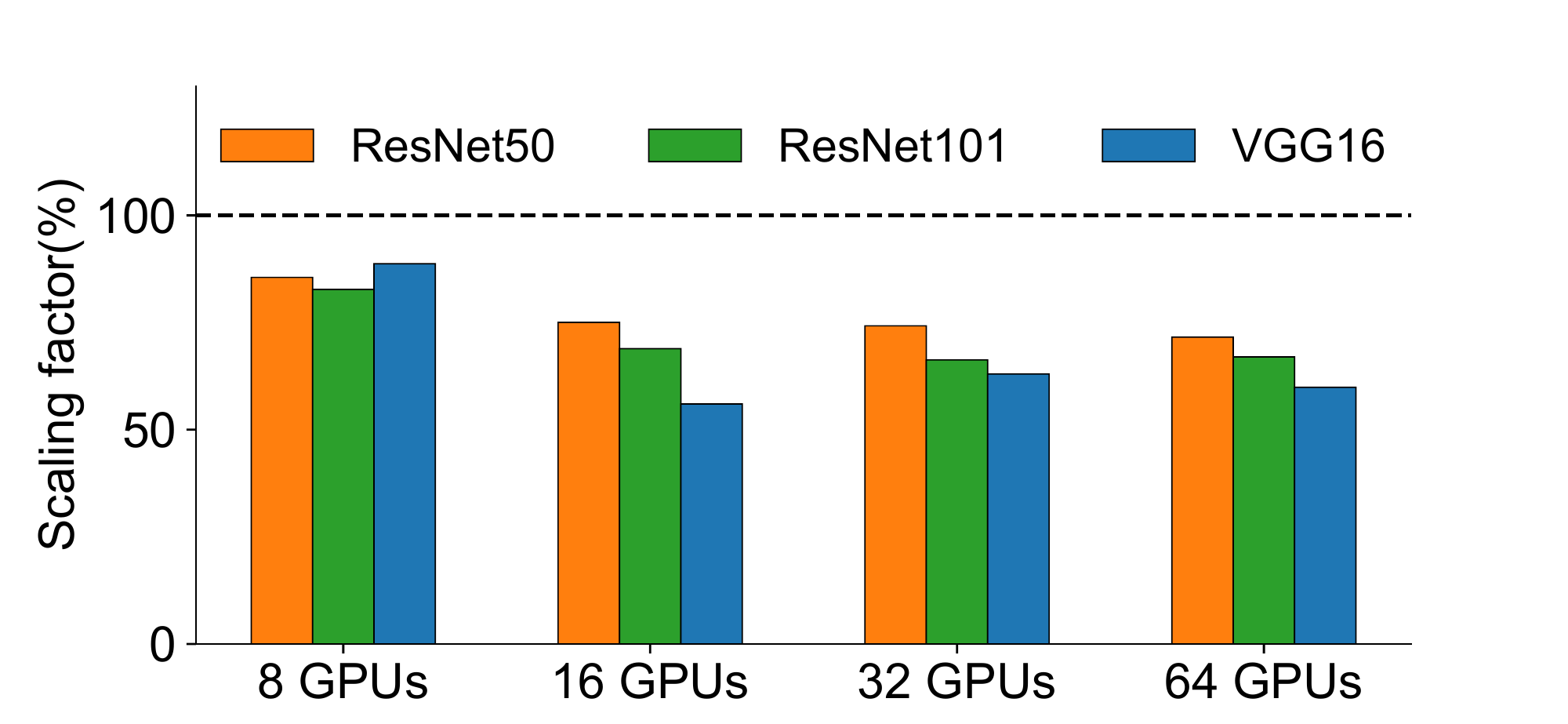}
    \vspace{-0.15in}
    \caption{Scaling factor vs. number of servers involved. }
    \medskip
    % \small The scaling factor is computed as following:
    % $f = \frac{Tput_{GPUCluster}}{n \times Tput_{oneGPU}}$, where $n$ is the
    % number of GPU involved in the GPU cluster.
    \label{fig:eval_scaling_vs_num_nodes}
    \vspace{-0.15in}
\end{figure}

\subsection{What is the current scaling factor?}

The first step is to understand the current scaling factor that can be
achieved by an off-the-shelf distributed training framework like Horovod. We use the throughput
of a single GPU (i.e., the number of images that can be processed by a GPU each
second) as the base throughput $T$. We vary the number of servers in the
experiments. For each case, we measure the total throughput that can be achieved
by the servers and compute the scaling factor based on
Equation~\ref{equ:scaling}. Figure~\ref{fig:eval_scaling_vs_num_nodes} shows the
scaling factor for each model under different numbers of servers. Remember that
we use p3dn.24xlarge instances, each of which contains 8 GPUs. So the figure
shows the scaling factor from 8 GPUs to 64 GPUs. The results indicate that the
scaling factors for ResNet50, ResNet101 and VGG16 are 75.05\%, 68.92\%, and
55.99\% for 2 servers, 74.24\%, 66.28\% and 63.01\% for 4 servers,
and 71.6\%, 66.99\% and 59.8\% for 8 servers. ResNet50
achieves better scaling factors than ResNet101 and VGG16 as it has a relatively
% smaller model size and a higher computation-to-communication ratio.
smaller model size to ease the communication burden.
Nevertheless, for all the three models, Horovod cannot achieve a scaling factor
of more than 76\% on AWS.

These results confirm that the current off-the-shelf
distributed training framework like Horovod cannot achieve linear scaling but with a significant gap.

\begin{figure}[t]
    \centering
    \includegraphics[width=0.9\linewidth]{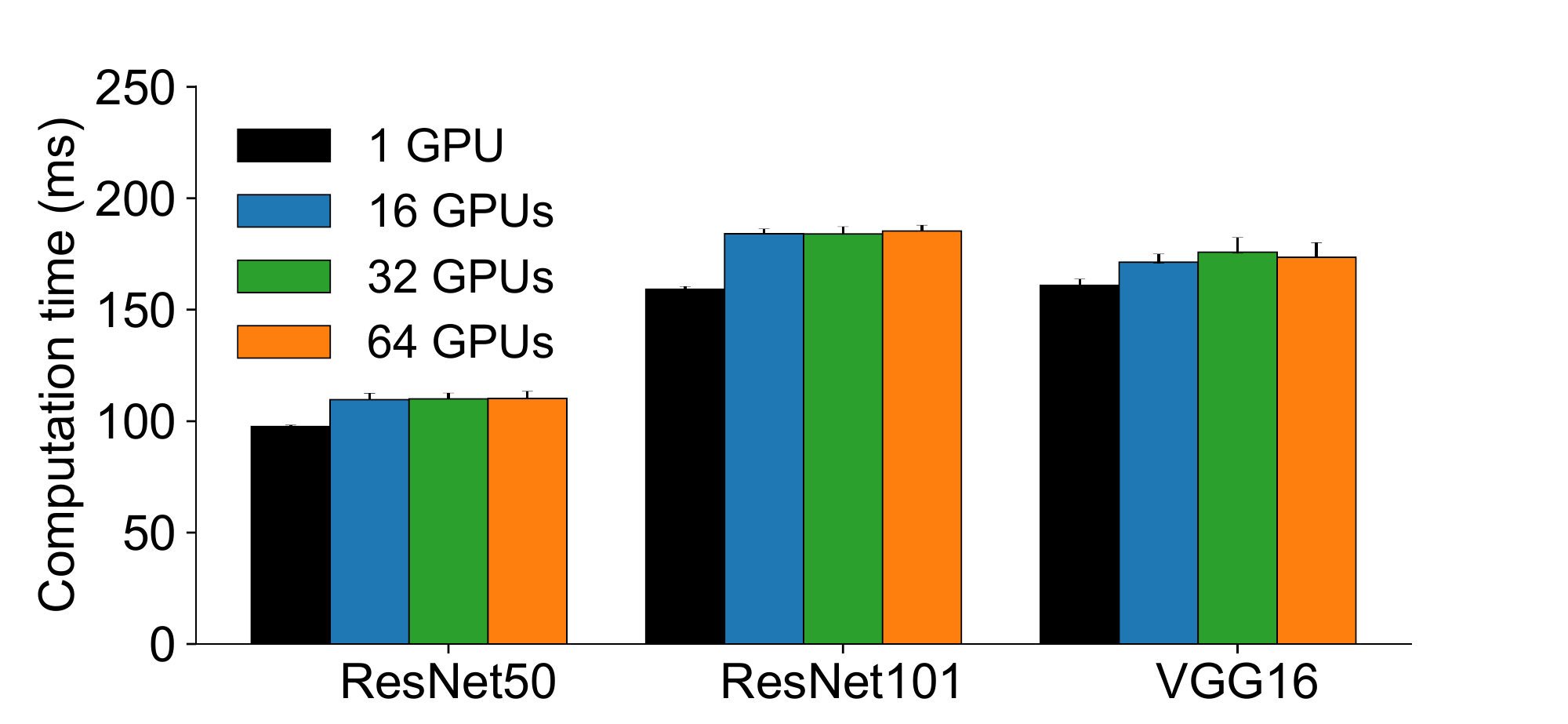}
      \vspace{-0.15in}
    \caption{Computation time vs. number of servers.}
    \vspace{-0.15in}
    \label{fig:eval_computation}
\end{figure}

\subsection{Is computation the bottleneck?}
Distributed training contains a computation component and a communication
component. To figure out why linear scaling cannot be achieved, we start with
the computation component. In the computation component, each worker feeds a
batch of labeled images to the neural network model and computes the gradients
locally. If the computation time for a worker to finish its batch increases with the number of workers, then the computation component would be the bottleneck of
distributed training.

Figure~\ref{fig:eval_computation} shows the computation time (for the forward and
backward pass) for the three models with different
number of workers. The computation time keeps almost the same, regardless of the
number of workers. The time gap between single GPU and multiple GPUs comes mainly from
two factors. 
%\haibin{Should we have a finer grained measure here, splitting the computation time for forward and backward? I suspect that the increase in computation time only happens during the backward pass, due to NCCL allreduce. NCCL allreduce introduces kernel launch overhead, and consumes some GPU register and memory transactions (I will try to get updated results once I have time.)}
First, the runtime for the backward pass in distributed training not only includes backward operations but also the all-reduce operations since they are asynchronous on GPU and overlapped. Whereas, for the single GPU case, there is no all-reduce operation. Second, Horovod injects a hook for each layer in the model during distributed training, which does not exist in single GPU training. 
% The first factor is due to
% the asynchronous and the pipeline nature of backward and all-reduce operations,
% which causes inaccurate measurement. 
%The first factor is due to facts that both backward operations and all-reduce operations
%are asynchronous on GPU and these operations are overlapping with each other. 
%Due to the asynchronous nature, 
%we need \texttt{torch.cuda.synchronize()} or \texttt{torch.cuda.Event.synchronize()}
%clause to make sure backward is completed. 
%During the backward phase, all-reduce process
%also starts working, e.g. after backward computation on last several layers, all-reduce operations on 
%those layers started.
%Thus, synchronization operations, e.g \texttt{torch.cuda.synchronize()}, also waits for launched
%all-reduce operations to be completed.
%Thus, we can see the time gap here is an overlook for computation time differences.
However, even considering this computation time gap as an inevitable side effect,
the scaling factor should still be bounded around 90\% instead of the measured 56\%-75\%, because the measured computation time increases at most 15\% in distributed training. 
Thus, we argue computation time difference here is not a factor for distributed
training not able to scale linearly.

% Since the network is running at 100 Gbps, we also want to see
% if the network speed affects the computation time. We use Linux Traffic Control
% to limit the bandwidth on each server, and measure the computation time. As
% shown in Figure~\ref{}, the computation time also does not change under
% different network speed.

% \yida{This paragraph is somewhat disconnected with the previous one.}The results are expected, because the computation component for distributed
% training is embarrassingly parallel. The task for each worker is entirely local
% and does not involve the network, thus the amount of work, i.e., processing an
% image batch, are independent to other workers. As a result, the time
% taken by the computation component stays the same with different number of
% workers and thus the computation component is not
% the bottleneck of distributed training.

\begin{figure}[t]
  \centering
  \includegraphics[width=\linewidth]{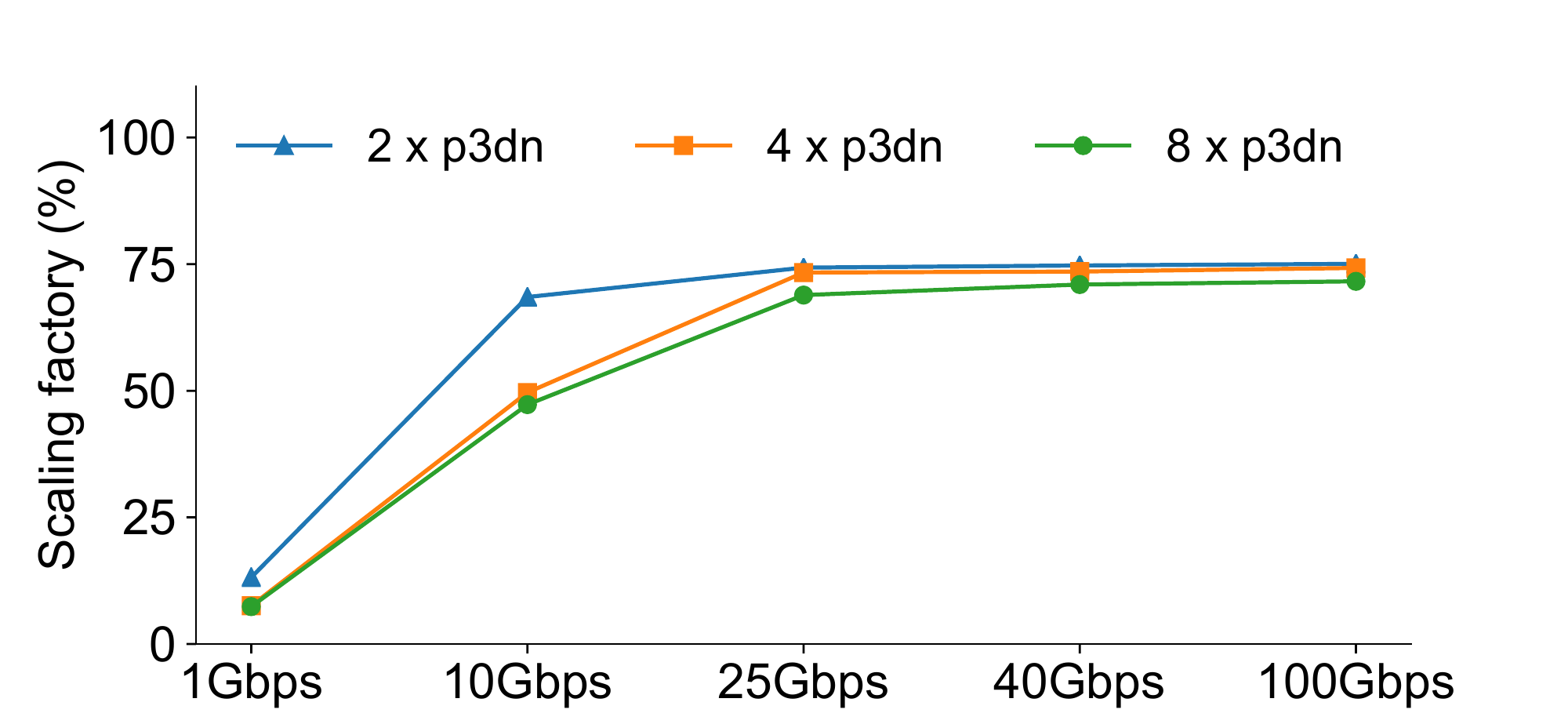}
    \vspace{-0.15in}
    \caption{Scaling factor change with bandwidth (ResNet50).}
    \vspace{-0.15in}
    \label{fig:eval_bw_affect_scaling_resnet50}
\end{figure}

\subsection{Is network the bottleneck?}

Now we turn to the communication component. Since the computation component takes the same amount of time regardless of the number of servers, then the only
possibility is that the communication component is the bottleneck when the
system scales out. To see whether this is the case, we first measure the scaling
factor with different network bandwidths. As shown in
Figure~\ref{fig:eval_bw_affect_scaling_resnet50}, the scaling factor for
ResNet50 does increase when the network bandwidth increases. In the case of two
servers, the scaling factor grows from 13\% to 68\% when the bandwidth increases
from 1 Gbps to 10 Gbps. This is understandable as with higher bandwidth, it
takes less time for the workers to exchange the same amount of data. The
scaling factor is lower with more workers as they have more data to exchange,
based on the all-reduce algorithm.

\begin{figure}[t]
    \centering
    \subfigure[Network (Recv) utilization.]{
        \label{fig:eval_network_recv_util}
        \includegraphics[width=0.9\linewidth]{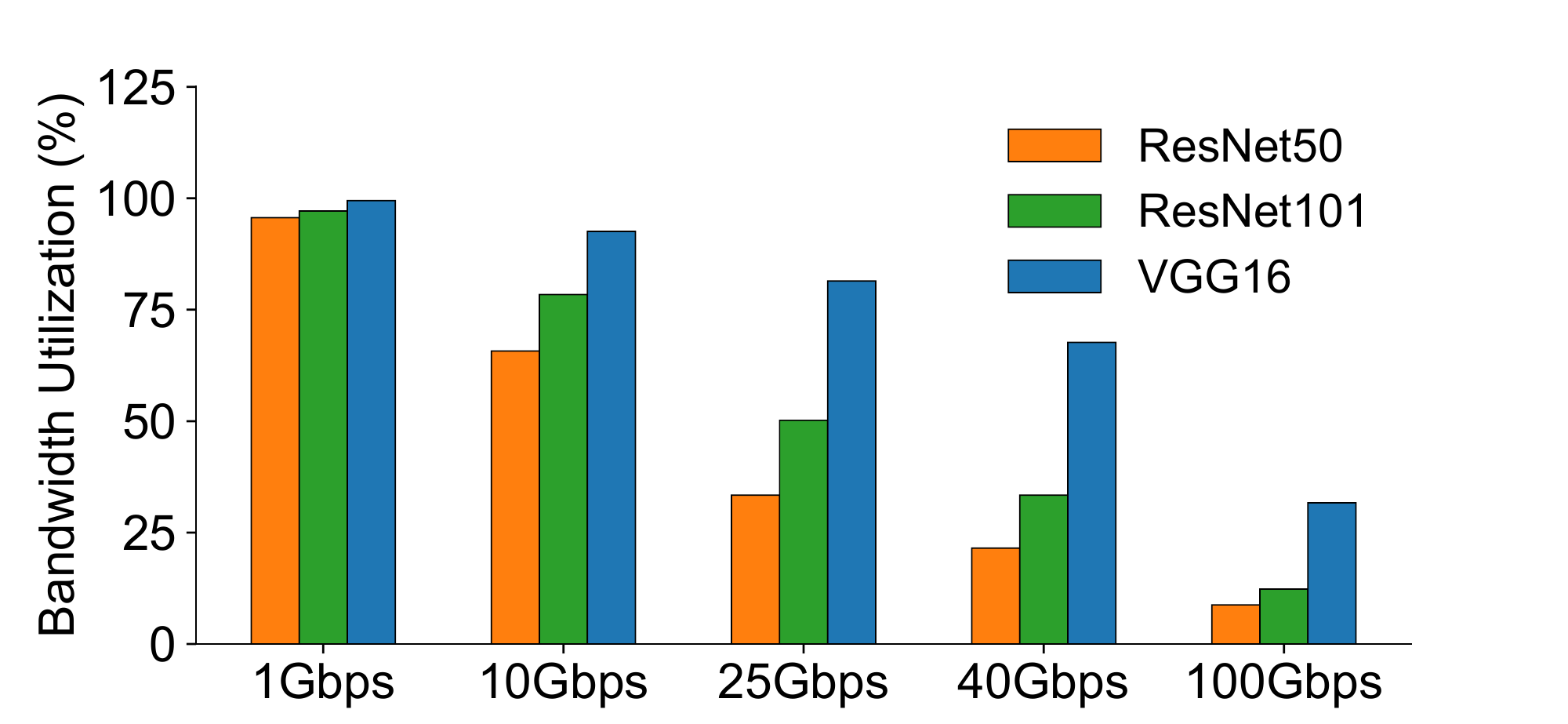}}
    \subfigure[Network (Send) utilization.]{
        \label{fig:eval_network_send_util}
        \includegraphics[width=0.9\linewidth]{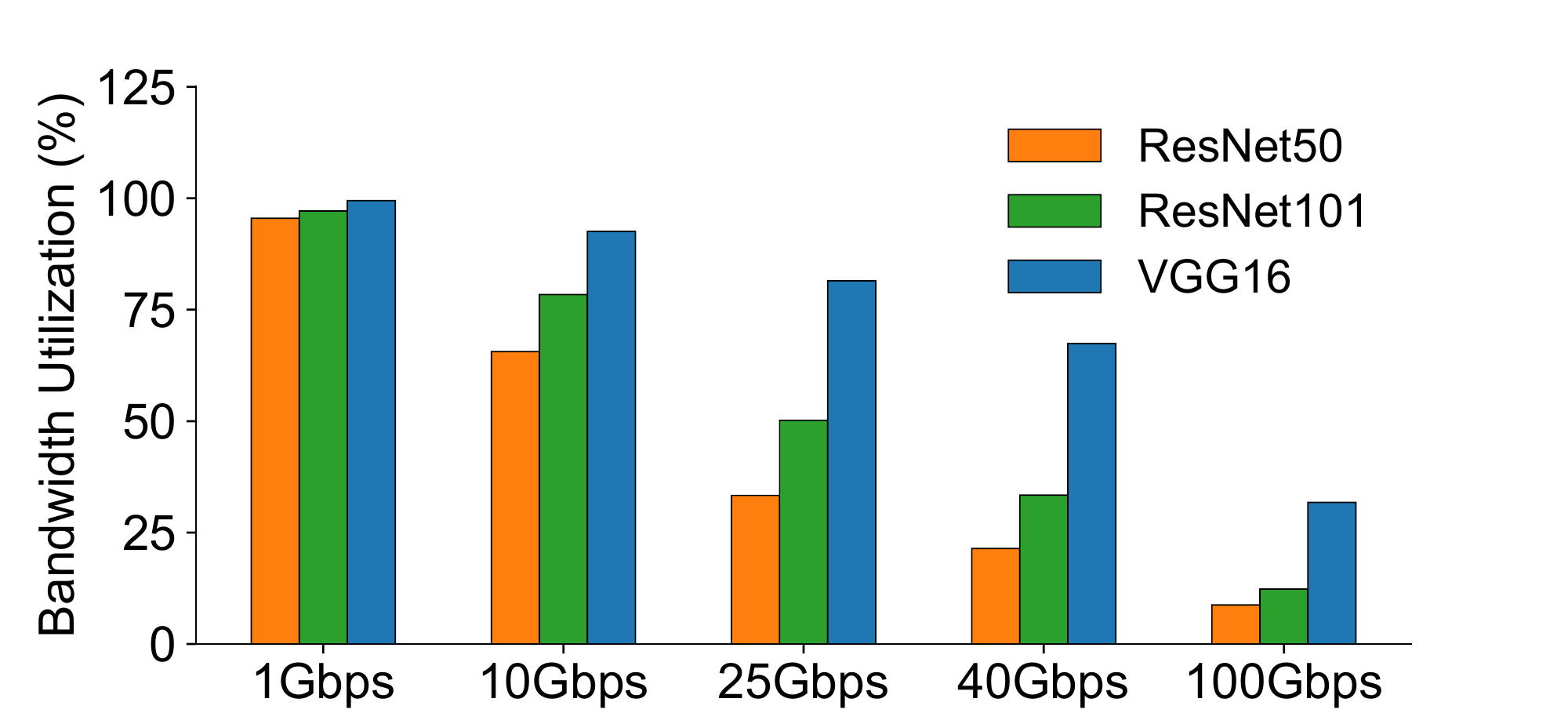}}
    \vspace{-0.15in}
    \caption{Network bandwidth utilization. }
    \vspace{-0.15in}
    \label{fig:eval_network_util}
\end{figure}

However, contrary to the common belief that the network is too slow to send the
gradients, Figure~\ref{fig:eval_bw_affect_scaling_resnet50} shows that the lines plateau
after 25 Gbps. This means the system can not
benefit from a faster network. To validate this, we measure the network
utilization of the servers by recording real time network throughput. Figure~\ref{fig:eval_network_util} indicates that the
servers do fully utilize the network at low bandwidth (e.g., 1 Gbps), but they
only use a small fraction of the bandwidth at high bandwidth (e.g., 100 Gbps).
This means merely adding bandwidth to make the network faster is not useful for
improving scaling factor after a certain point.

\begin{figure}[t]
    \centering
    \includegraphics[width=0.9\linewidth]{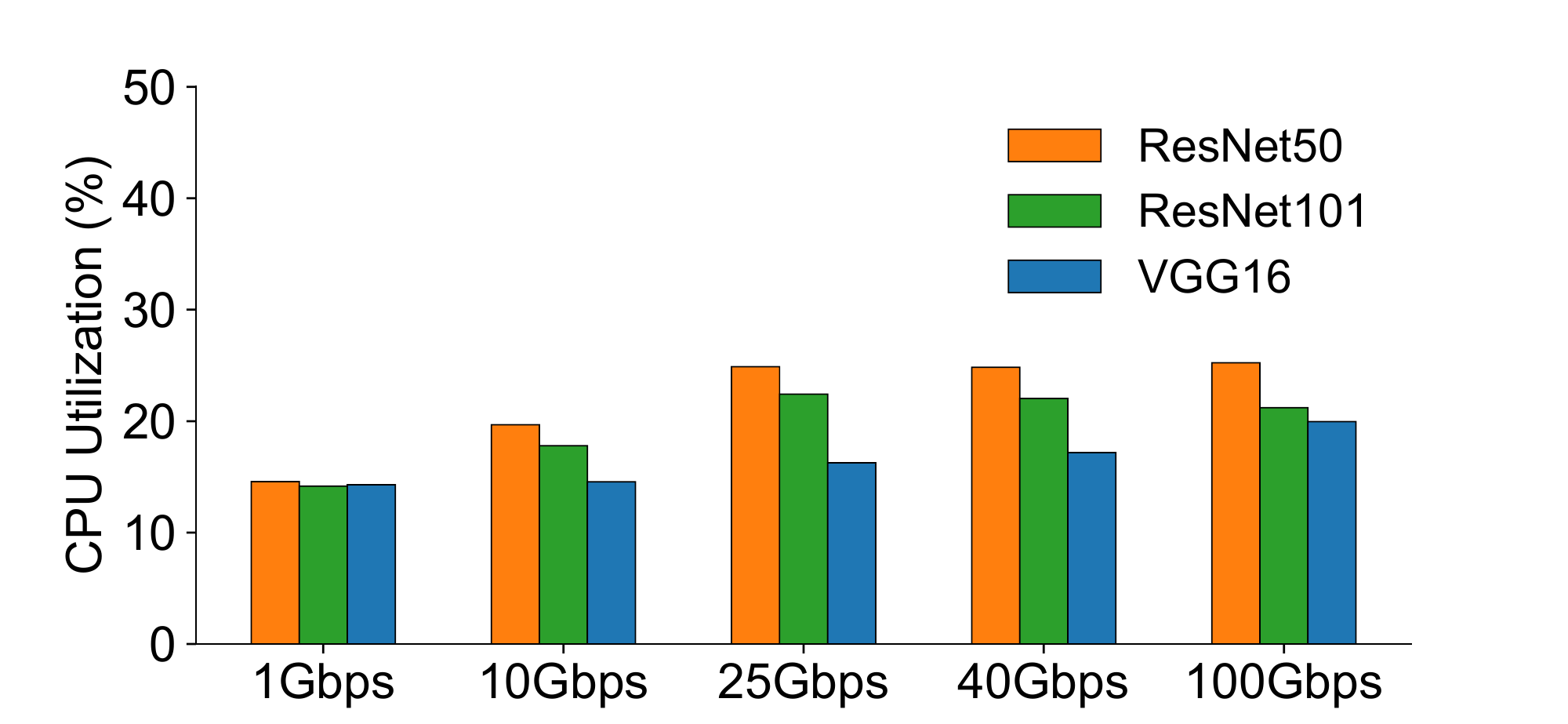}
    \vspace{-0.15in}
    \caption{CPU utilization.}
    \vspace{-0.15in}
    \label{fig:eval_cpu_util}
\end{figure}

\begin{figure*}[t]
  \centering
  \subfigure[ResNet50.]{
    \label{fig:eval_bw_sim_scaling_resnet50}
    \includegraphics[width=0.32\textwidth]{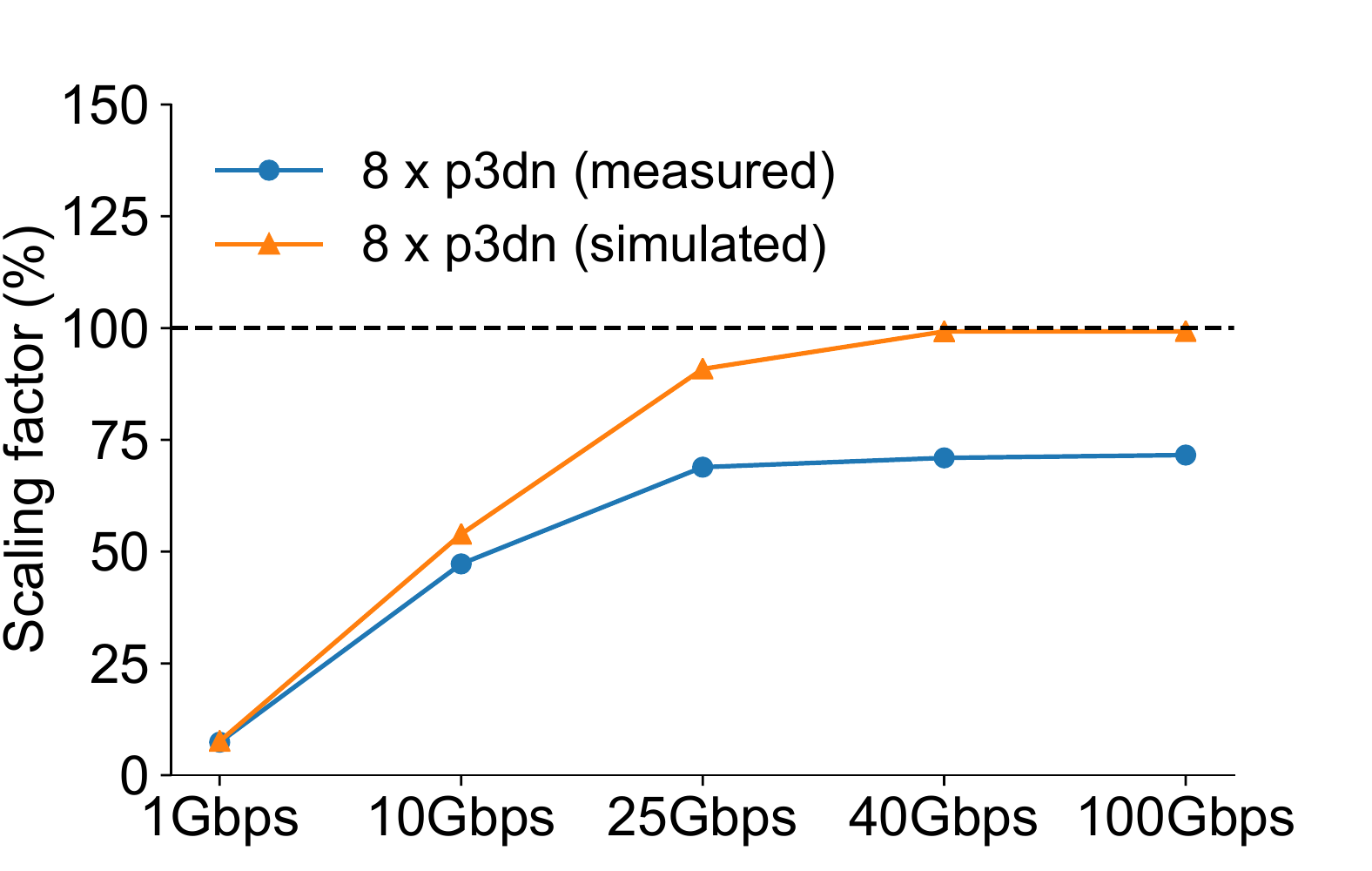}}
  \subfigure[ResNet101.]{
    \label{fig:eval_bw_sim_scaling_resnet101}
    \includegraphics[width=0.32\textwidth]{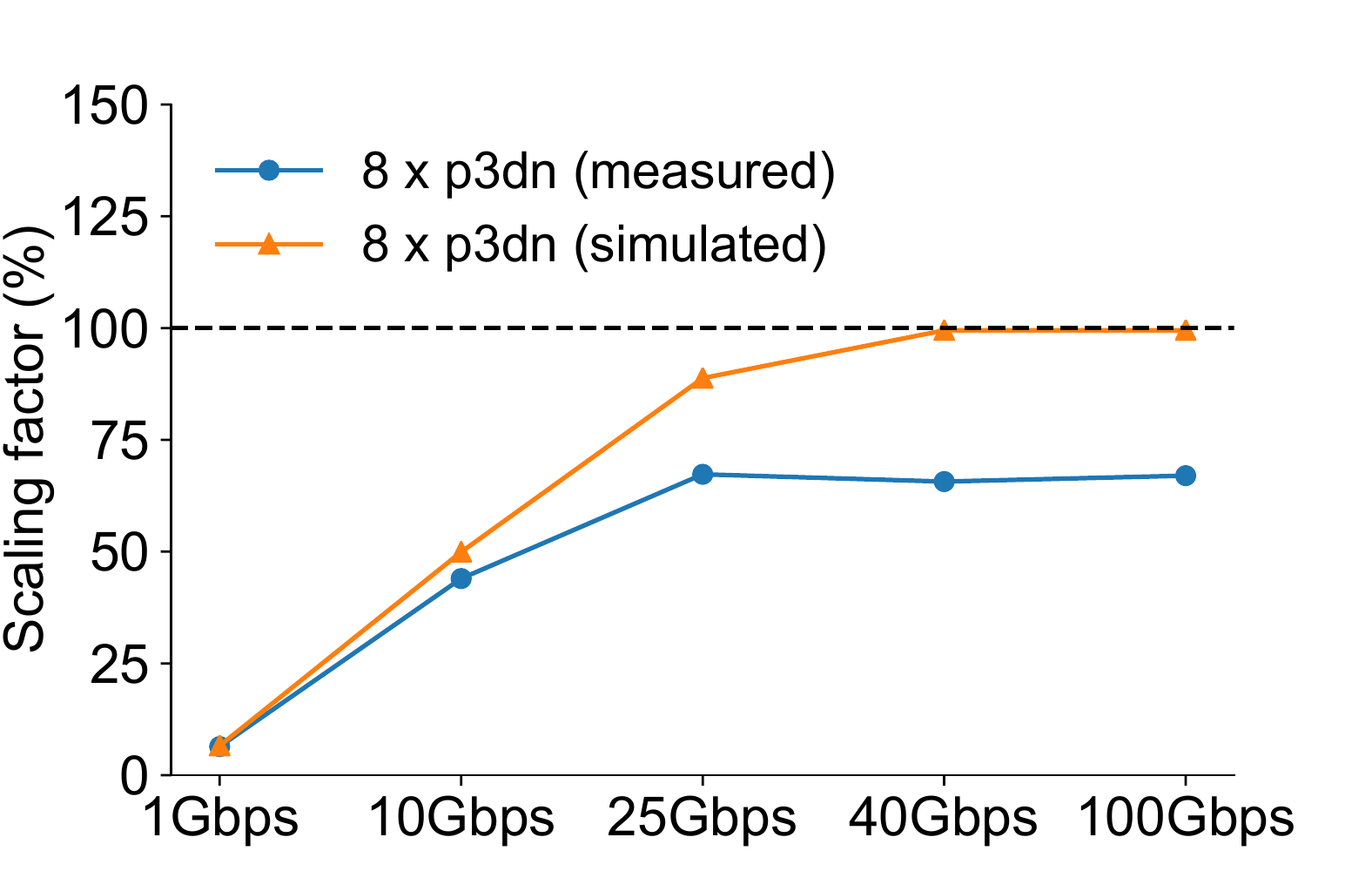}}
  \subfigure[VGG16.]{
    \label{fig:eval_bw_sim_scaling_vgg16}
    \includegraphics[width=0.32\textwidth]{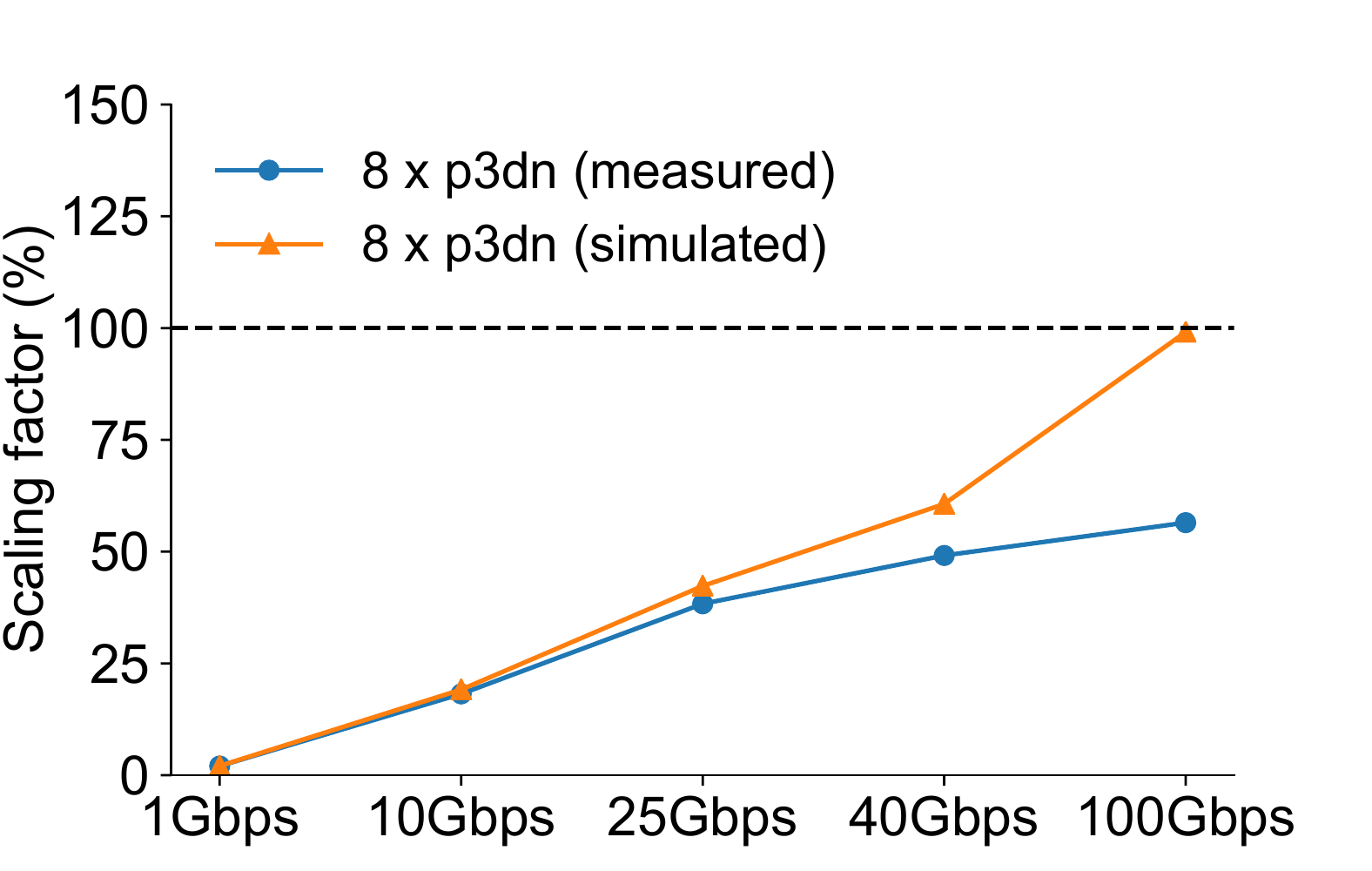}}
  \vspace{-0.15in}
  \caption{Simulated scaling factor vs measured scaling factor in different bandwidth.}
  %\yida{The behavior of VGG16 simulation is different from the other two (not growing slower and slower), we probably should explain that a bit.
%   (main points here are that 1)simulated results are first tight with measured results, 2) we can finally get close 100\%. 
%   I feel it is due to the large model size of VGG16, so it should get slower close to 95 Gbps. but it is not clear, because we don't have data to support.)}
%   \yida{I get your points, however, the question remains. I am not comfortable without an explanation here. We are doing the simulation, everything should be under control.}
% I think we are going to study the bandwidth effects on next paper in more details. 

  \vspace{-0.15in}
  \label{fig:eval_bw_sim_scaling}
\end{figure*}

One possibility for low utilization at high bandwidth is that the CPU might be
the bottleneck, as the experiments run Horovod over TCP and it is known that
running TCP at high speed like 100 Gbps is CPU-intensive. However, the
computation of distributed training is mostly done by GPUs, and most GPU
instances are equipped with sufficient amount of CPUs (e.g., 96 vCPUs in a
p3dn instance used by our experiments). 
Figure~\ref{fig:eval_cpu_util} shows the CPU
utilizations while training three models on eight p3dn instances under five different
network speeds. It confirms that the CPU utilization is low, and thus CPU is not
the bottleneck for saturating 100 Gbps network bandwidth.

In conclusion, the measurement confirms that the communication component is the
bottleneck. But contrary to the common belief, it is not because the network is
too slow to send data. The root cause is the poor implementation of the network
transport that cannot fully utilize the available bandwidth for the
communication component.

%%% Local Variables:
%%% mode: latex
%%% TeX-master: "../paper"
%%% End:

\section{What-If Analysis}

Given the low network utilization, a natural question is what if the network can
be fully utilized. In this section, we perform a what-if analysis to evaluate
the scaling factor under full network utilization. Given the promise of many
proposals on application-layer optimizations, we also use what-if analysis to
show what additional improvements these proposals can bring if the network is
fully utilized.

\subsection{What if network can be fully utilized?}
\label{whatif-net}

We first perform a what-if analysis to see what scaling factors can be achieved
if the network is fully utilized. To do the what-if analysis, we need detailed
logging information first, then perform simulation based on the timing logs.
 We take the white-box approach to directly add logging code to
training scripts to retrieve detailed timing information for what-if analysis.
Specifically, we add hooks for parameters in the model to get the 
\emph{gradient-computation-done} time for different layers of the model. 

For the simulation, we have two
processes, \emph{backward process} and \emph{all-reduce process}. 
Two processes communicate through a message queue.
The backward process simulates the backward computation which is based on the timing log
of \emph{gradient-computation-done}.
% once gradients of a certain layer is computed
% it notifies the \emph{all-reduce process} for all-reduce operation by sending the data to 
% a \emph{queue}. 
% The \emph{backward process} continues backward pass without waiting for \emph{all-reduce process}.
The \emph{backward process} does not request 
\emph{all-reduce process} right after backward computation done for a certain layer.
Instead, it buffers gradients of several layers for all-reduce.
We use a heuristic buffering strategy, which refers to Horovod fusion buffer~\cite{sergeev2018horovod}. 
Specifically, the \emph{backward process} has a timeout window of \emph{5 ms} and a gradients buffer size of \emph{64 MB} for
batching gradients for the all-reduce operations.
Once the timeout criterion or buffer size
limit is satisfied, it notifies the all-reduce process for all-reduce operation.
The \emph{all-reduce process} uses Reduce-scatter with Allgather procedures to
complete all-reduce operation. 
% The all-reduce process also uses 
% sleep operation to mimic the cost of network transitions and vector additions.
The transition time is computed as $(2 S
 (N -1) / N) / bw $, where $N$ is the number of workers/GPUs involved, $S$ is the size
for all-reduce and $bw$ is the network bandwidth.
The cost of vector additions is estimated as $(N - 1) \times AddEst(S /
N)$, where $AddEst(x)$ is the function for estimating the time of element-wise
adding of two vectors in the size $x$.
To fairly estimate the vector addition time cost, we first empirically evaluate
time cost of vector-add with various vector sizes on V100 GPU, and then use linear
interpolation to get $AddEst(x)$.

To get the scaling factor for a certain bandwidth and number of workers,
we start \emph{backward process} and \emph{all-reduce process} 
at the same time for simulation.
For each data batch, we denote the time for all-reduce process to complete 
as $t_{sync}$,
and denote the time for backward computation 
$t_{back}$, thus we can get the overhead for all-reduce operation as
$t_{overhead}=t_{sync}-t_{back}$. Then, we can use processing time of a batch data on single GPU, $t_{batch}$, 
to get the simulated scaling factor: $ f_{sim} = t_{batch} / (t_{batch}+t_{overhead})$.

\begin{figure}[t]
  \centering
  \includegraphics[width=0.9\linewidth]{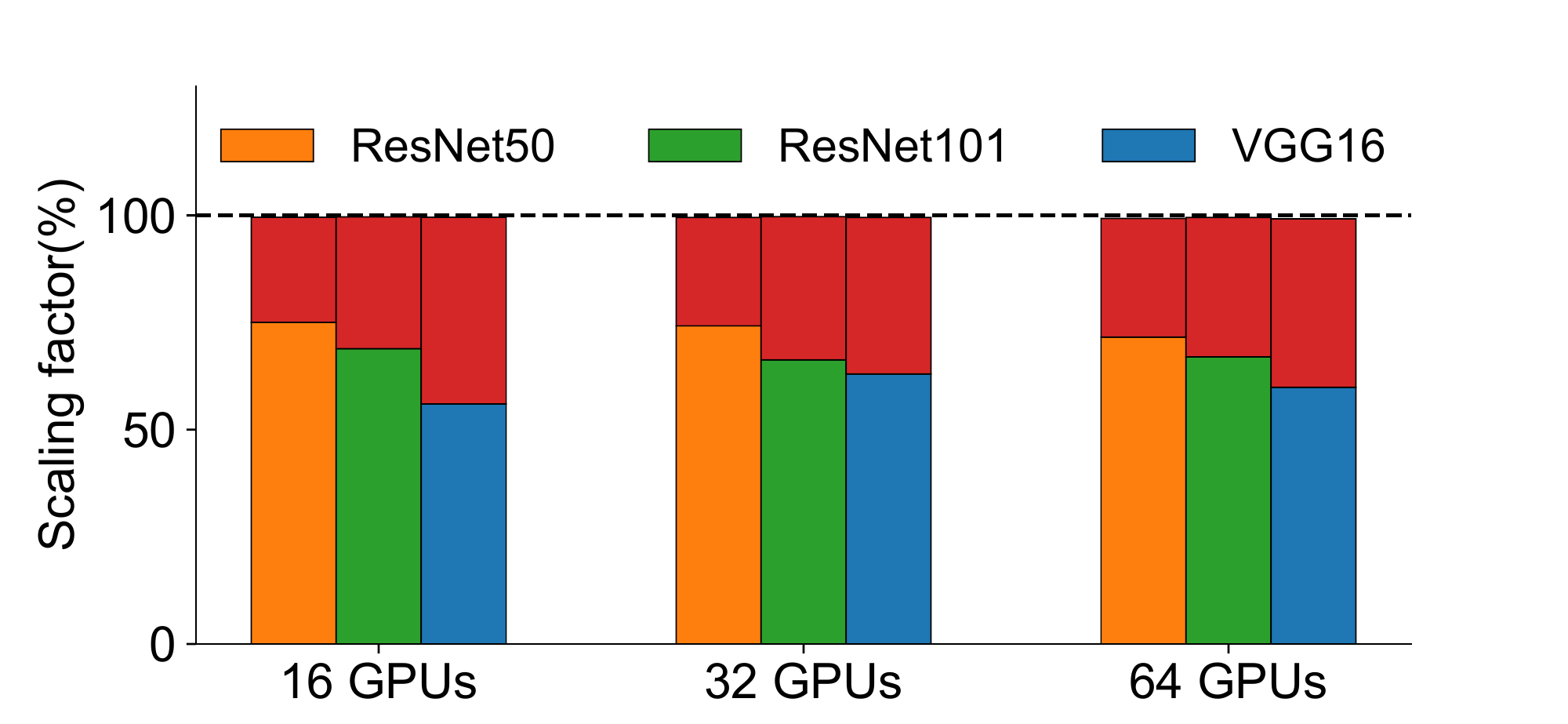}
    \vspace{-0.15in}
    \caption{Simulated scaling factor under 100 Gbps network. Red parts denote
      the gap to simulated results.}
    \vspace{-0.15in}
    \label{fig:eval_simulated_vs_num_nodes}
\end{figure}

Figure~\ref{fig:eval_bw_sim_scaling} shows the scaling factors of the three models under different
network speeds assuming the network is fully utilized, and compares them with
the scaling factors actually achieved by Horovod. We can see that under low
network speeds (i.e., 1 Gbps and 10 Gbps), the two lines are very close. This
confirms the results in Figure~\ref{fig:eval_network_util} that the network is fully utilized under
low speeds, and also validates the correctness of the what-if simulator. Under
high network speeds (i.e., after 25 Gbps), the two lines begin to diverge
significantly. While the system can theoretically achieve close to 100\% scaling
factor under 100 Gbps for ResNet50, ResNet101 and VGG16,
in practice it only achieves 75\%, 67\% and 60\%, respectively. 

We also use the what-if analysis to evaluate the scaling factors under
different numbers of workers assuming that the network is fully utilized. The results
are shown in Figure~\ref{fig:eval_simulated_vs_num_nodes}. Again, we see all
of three models can
achieve close to 100\% scaling factors when the network is fully utilized even
for 64 GPUs. 
Overall, the what-if analysis confirms that distributed training can benefit
from high network bandwidth, moreover the scaling factor can be improved to near 100\% if the
network is fully utilized.

\begin{figure}[t]
%  \vspace{-0.15in}
  \centering
  \subfigure[Compression in 10Gbps network.]{
    \label{fig:eval_compression_10G}
    \includegraphics[width=0.45\textwidth]{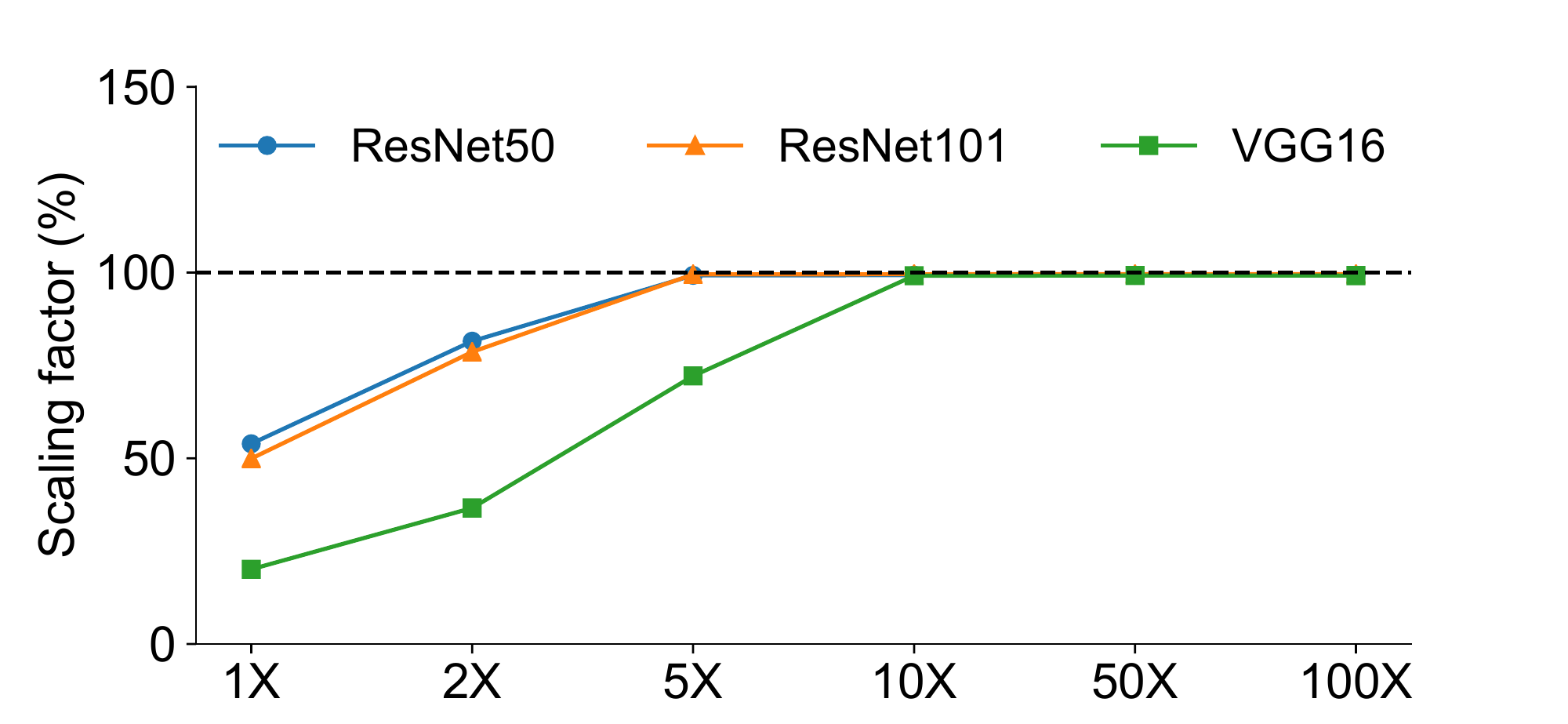}
    % \vspace{0.1in}
  }
  \subfigure[Compression in 100Gbps network.]{
    \label{fig:eval_compression_100G}
    \includegraphics[width=0.45\textwidth]{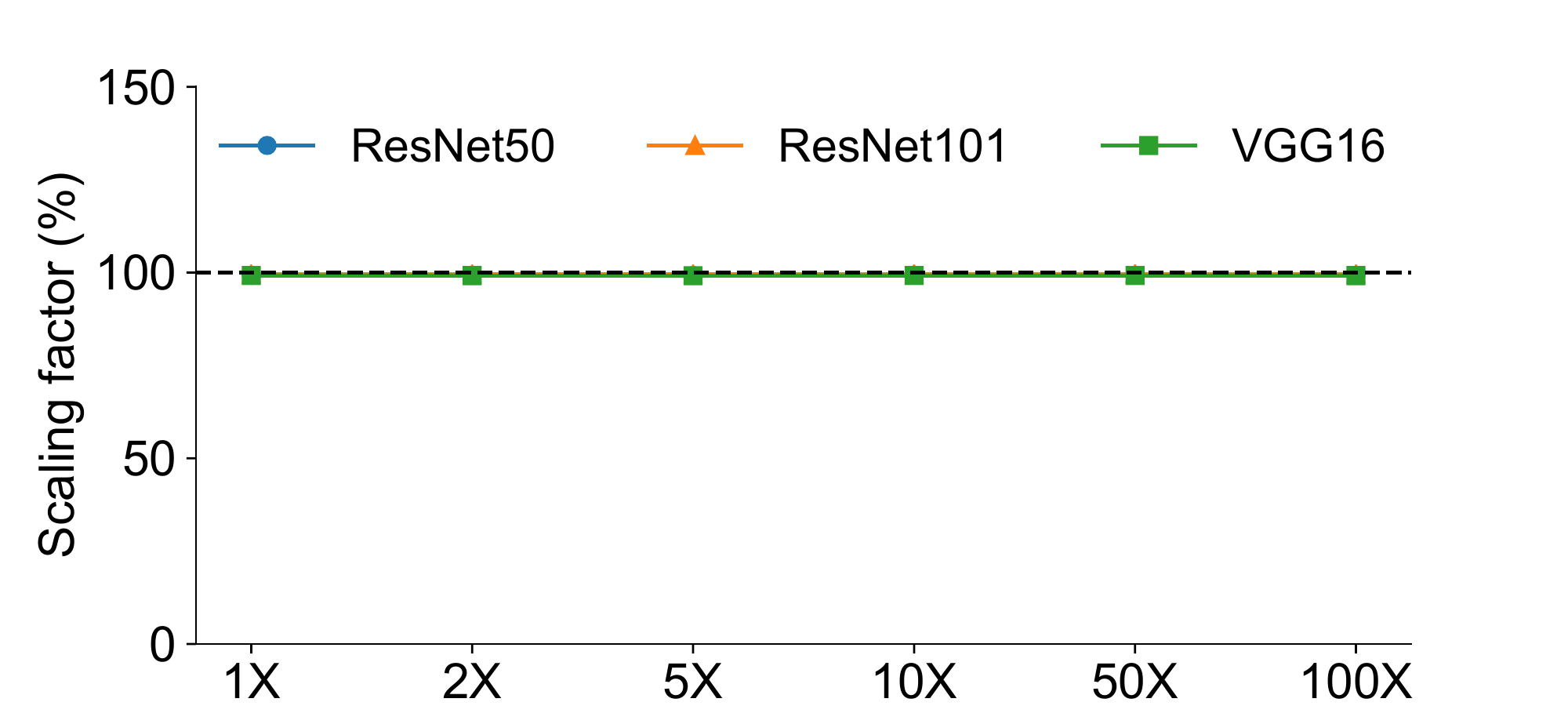}}
%   \vspace{-0.15in}
  \caption{Simulated scaling factor under different compression ratio. } 
%   \vspace{-0.15in}
  \label{fig:eval_compression}
\end{figure}

\subsection{How useful are application-layer optimizations?}

In this section our analysis targets a well-studied application-level
optimization technique---gradient compression. We keep other simulation
steps the same as we do in \S\ref{whatif-net}, but divide the
time cost of gradients transmission by the compression ratio we choose.
We use this setup for the simplicity. As one would imagine, the compression
could possibly reduce the vector-add cost (e.g. half-precision
vector-add, top-k percent gradients for all-reduce) to further boost the
simulated scaling factor. But as shown in Figure \ref{fig:eval_compression},
the simplified simulation is good enough to justify the claim we want to
make, which is we probably would not need that high compression ratio as advocated
in past works~\cite{lin2017deep, mishchenko2019distributed, lim20183lc}. The compression ratio 10$\times$ is large enough for models
like VGG16 to get scaling factor near 100\% in 10 Gbps network, which is commonly
available at cloud platform like AWS, GCloud and Azure. As a comparison, the results in 100 Gbps are also reported to
indicate that compression is not that useful in high-speed networks, which is the
typical network configuration for high-end GPU servers like aws-p3dn. In
conclusion, gradient compression techniques are useful in low-speed networks, but it is
not necessary to have a large compression ratio in contemporary network environments.

% So, are other solutions useful?

% \begin{itemize}
% \item Scheduling: Scaling factor vs. network speed, assuming network is highly
%   utilized
% \item Compression: Scaling factor vs. network speed, assuming network is
%   highly utilized
% \item In network aggregation: Scaling factor vs. network speed, assuming network is highly utilized

% \end{itemize}

% Conclusion they can improve at most 10--20\%

\iffalse
\begin{table}[t]
\centering
\small
\begin{tabular}{l r r r r r}
\toprule
 & \textbf{1 Gbps} & \textbf{10 Gbps} & \textbf{25 Gbps} & \textbf{40 Gbps} & \textbf{100 Gbps} \\
\midrule
ResNet50 & 1.22\% & 6.62\% & \textbf{20.26\%} & 6.88\% & 0.99\% \\
\bottomrule
\end{tabular}
\vspace{-0.05in}
\caption{Improved Scaling factor of ByteScheduler. }
\vspace{-0.05in}
\end{table}
\fi

%%% Local Variables:
%%% mode: latex
%%% TeX-master: "../paper"
%%% End:

\section{Discussion and Future Work}

\paraf{Rationale behind the findings.} At first glance, our findings may be
surprising, indicating that the scaling factor can be close to
100\% if the network is fully utilized. 
These findings, however, are quite reasonable because
of two important factors. First, the network runs at high speed. Under 100 Gbps,
it only takes 7.8 ms, 13.6 ms and 42.2 ms to transmit all parameters of ResNet50, ResNet101 and
VGG16, respectively. Second, there is a significant overlapping between
computation and communication. The all-reduce for the last layer can start as
soon as the backward process has computed the gradients of the last layer,
without waiting for the entire backward process to finish. This overlap is critical.
%The overlapping is
% critical for the scaling factor to achieve near 100\%, because otherwise the computation
% and communication would be sequential
In conclusion, combining the efficient communication
and the overlapping of computation and
communication, the scaling factor can achieve near 100\%.

\parabf{Generality of the results.} One essential question is how general are 
the results. The results are based on three models (ResNet50, ResNet101 and VGG16),
one particular device (NVIDIA V100), and one training strategy (all-reduce). As
part of our future work, we plan to expand the measurement and analysis to more
models (e.g., RNN-like sequence models and BERT), more
devices (different GPUs and other specialized processors), and more
training strategies (e.g., parameter server and asynchronous training). While the
actual numbers might differ, we expect that the conclusion would stay the same. 
i.e.,
because of high-speed networks and the intrinsic overlap between computation
and communication, increasing the network utilization would result in
almost linear scaling.
% significant improvements to the scaling factor for distributed training.

\parabf{Trade-off of application-layer optimizations.} The what-if analysis
indicates that gradient compression in the application layer only provides
meaningful improvements at low network bandwidths. 
% As such, it is useful for
% scenarios like federate learning in bandwidth-constraint settings, e.g.,
% multiple geographically-distributed organizations train a model over the wide
% area network (WAN) or multiple mobile devices train a model collectively over
% the cellular network.
% As the gradient compression does not necessarily help for distributed training with high-speed networks, 
We argue that it is not particularly useful for distributed
training on the cloud or an on-premises cluster equipped with GPUs or TPUs. Those machines are typically connected with
high-speed networks to fully utilize the processors. It does not make sense to build a cluster for distributed training with expensive specialized processors but a cheap, slow
network.

\parabf{The proper metric for scaling.} We use throughput to compute the scaling
factor. Another proper metric is to use the convergence time, i.e., the time to
train a model to reach a certain accuracy threshold. Ideally, with $n$ servers,
the convergence time should be cut by $n$ times (i.e., 100\% scaling factor).
This metric might be the most important metric cared by researchers and developers. 
We emphasize
that network-layer optimizations provide consistent performance on both metrics,
as it reduces the time to finish one iteration without changing the
number of iterations needed to reach a certain accuracy. Also, network optimizations are orthogonal to other techniques to accelerate the training process~\cite{you2018imagenet}. Gradient compression,
on the other hand, loses gradient information due to lossy compression, and can
prolong the convergence time, hurt the accuracy, and
even end up not being able to converge.

\parabf{What-if analysis for other approaches.} Besides gradient compression,
there are other application-layer and system-layer optimizations. For example,
% PipeDream~\cite{narayanan2019pipedream} uses model parallelism to reduce the amount of data that needs
% to be exchanged between different workers; % PipeDream actually increase the data due to the model parallelism 
 ByteScheduler~\cite{peng2019generic} orders the
gradient transmission of different layers to better overlap with forward computation;
and SwitchML~\cite{sapio2019scaling} uses a programmable switch to aggregate gradients and
reduce the communication size. These proposals all suggest significant reduction
on the training time. However, they are all compared to an off-the-shelf
distributed training framework like Horovod, which has a poor network transport
implementation. It would be
interesting to apply the what-if analysis to evaluate what additional
improvements they can provide if the network can be highly utilized.

\parabf{High-performance network transport for distributed training.} 
There is always an arms race between compute and network. When compute is improved, network becomes the bottleneck.
Our findings indicate that for today's
distributed training systems, the network speed is not a problem, but the
network transport implementation for the communication component is.
% Application-layer and system-layer optimizations have several drawbacks. They
% require changes to the applications or the training framework, and sometimes
% even cause model performance degradation. 
Compared to application-layer optimizations, e.g. gradient compression, network-layer optimizations do not trade training time off against model accuracy, 
and should be the first-order optimizations. As such, our results
are a call for the network community to develop high-performance network
transport to fully utilize modern high-speed networks
and to achieve linear scale-out.
Recently, AWS has provided Elastic Fabric Adapter (EFA) as an efficient network interface to bypass OS kernel for high-performance communication~\cite{EFA}, and achieved some encouraging scalability results by carefully tuning the training process~\cite{EFAtraining}. Developing high-performance network transport with kernel-bypass technologies in the context of distributed training is an interesting direction of future work.

\parabf{Acknowledgments.} We thank the anonymous reviewers for their valuable feedback. This work is supported in part by NSF grants CRII-1755646, CNS-1813487 and CCF-1918757, and an AWS Machine Learning Research Award. RA would like to acknowledge support from NSF under the BIGDATA awards IIS-1546482 and IIS-1838139, and NSF CAREER award IIS-1943251.

%%% Local Variables:
%%% mode: latex
%%% TeX-master: "../paper"
%%% End:

% \input{compTo}

% \section*{Acknowledgement}
% We would like to express appreciation to Chaokun Chang for 
% running experiments for timing log gathering. 

% \newpage
{
\def\UrlBreaks{\do\/\do-}
\bibliographystyle{ieeetr}

\bibliography{xin.bib}}

\end{document}